\documentclass[11pt, showpacs]{revtex4}
\usepackage{graphicx, epsfig}

\def\half{\frac{1}{2}}
\def\beq{\begin{equation}}
\def\eeq{\end{equation}}
\def\beqa{\begin{eqnarray}}
\def\eeqa{\end{eqnarray}}
\def\iar{\begin{array}{l}}
\def\ear{\end{array}}

\begin{document}

\title{The quantum entanglement contained in density matrices}
\author{Yong Zhou}
\affiliation{China University of Petroleum, College of Physics Science and Technology, 739 Bei'yi Road, Dongying Shandong 257061, China}

\begin{abstract}
We point out that density matrices can only be used to describe quantum states, so the entanglement contained in a density matrix is just quantum entanglement. This means a bipartite state described by a density matrix contains quantum entanglement, unless the density matrix has the form $\rho_{AB}=\rho_A\otimes \rho_B$. 
\end{abstract}

\pacs{03.67.Mn, 03.65.Ud} 
\maketitle

Quantum entanglement \cite{c1, c2, c3} is a very important phenomenon. It has important application in quantum teleportation \cite{c4, c5}, quantum cryptographic schemes \cite{c6}, dense coding \cite{c7} and quantum parallel computation \cite{c8}. The quantifying of the quantum entanglement degree of bipartite pure states has been well done \cite{c9}, but how to define and quantify general system's quantum entanglement degree is still an unresolved problem \cite{c10}.

It's known that density matrices are the concept of quantum mechanics, so it can only be used to describe quantum states. Thereore we can deduce that the entanglement contained in a density matrix is just quantum entanglement. This means that only the density matrix having the form $\rho_{AB}=\rho_A\otimes \rho_B$ doesn't contain quantum entanglement. Any other density matrix, including the following one \cite{c11}:
\beq
  \rho_{AB}\,=\,\sum_{i=1}^{n}p_i\,\rho^A_i\otimes\rho^B_i\,,
\eeq
where $n>1$, contains quantum entanglement.

Obviously this conclusion contradicts the point of a number of scientists who considered the entanglement in Eq.(1) is classical entanglement \cite{c10, c11, c12}. Here we present two proofs to prove our point. Firstly consider the three systems: A, B and C, if A gets quantum entangled with BC group in a pure state:
\beq
  \vert\psi_{ABC}\rangle\,=\,\sum_i\sqrt{p_i}\vert A_i B_i C_i \rangle\,,
\eeq
where $\{ \vert A_i\rangle \}$, $\{ \vert B_i\rangle \}$ and $\{ \vert C_i\rangle \}$ are the orthonormal bases of the corresponding systems, the reduced density matrices describing A and B group and A and C group are:
\beq
  \rho_{AB}\,=\,\sum_i p_i \vert A_i B_i \rangle\langle A_i B_i\vert\,, \hspace{10mm}
  \rho_{AC}\,=\,\sum_i p_i \vert A_i C_i \rangle\langle A_i C_i\vert\,.
\eeq
If the point of the literatures \cite{c10, c11, c12} is right, there isn't quantum entanglement between A and B systems and A and C systems. We find this conclusion contradicts a logical fact: if A gets entangled with BC group, A must get entangled with at least one of B and C systems, otherwise A will get not entangled with BC group. Obviously only accepting the point that Eq.(1) contains quantum entanglement can resolve this problem.

Another evidence can be found in quantum teleportation. Suppose B system is far away from A system, A and C systems are in a same location, and A, B and C systems are in the quantum states:
\beq
  \rho_{AB}\,=\,\half(\vert A_1 B_1\rangle \langle A_1 B_1\vert +\vert A_2 B_2 \rangle \langle A_2 B_2\vert )\,,
  \hspace{10mm} \vert \psi_C\rangle\,=\, c_1\vert C_1\rangle + c_2\vert C_2\rangle\,.
\eeq
Perform an union measurement for A and C systems,we will obtain one of the following possible states:
\beqa
  \vert \psi_{AC1}\rangle\,=\, \frac{1}{\sqrt{2}}(\vert A_1 C_1\rangle + \vert A_2 C_2\rangle)\,, \hspace{10mm}
  \vert \psi_{AC2}\rangle\,=\, \frac{1}{\sqrt{2}}(\vert A_1 C_1\rangle -  \vert A_2 C_2\rangle)\,, \nonumber \\
  \vert \psi_{AC3}\rangle\,=\, \frac{1}{\sqrt{2}}(\vert A_1 C_2\rangle + \vert A_2 C_1\rangle)\,, \hspace{10mm}
  \vert \psi_{AC4}\rangle\,=\, \frac{1}{\sqrt{2}}(\vert A_1 C_2\rangle -  \vert A_2 C_1\rangle)\,,
\eeqa
accordingly B system will change its state to one of the following possible states:
\beqa
  \rho_{B1}\,=\,\vert c_1\vert^2 \vert B_1\rangle\langle B_1\vert +\vert c_2\vert^2 \vert B_2\rangle\langle B_2\vert\,,
  \hspace{10mm} \rho_{B2}\,=\,\rho_{B1}\,, \nonumber \\
  \rho_{B3}\,=\,\vert c_2\vert^2 \vert B_1\rangle\langle B_1\vert +\vert c_1\vert^2 \vert B_2\rangle\langle B_2\vert\,,
  \hspace{10mm} \rho_{B4}\,=\,\rho_{B3}\,.
\eeqa
It is obvious that B final state contains part of the information of C initial state: the module squares of the expanding coefficients of C initial state. On the other hand, if use the accepted quantum entangled state
\beq
  \vert \psi_{AB} \rangle\,=\,\frac{1}{\sqrt{2}}(\vert A_1 B_1\rangle + \vert A_2 B_2 \rangle)\,,
\eeq
to replace $\rho_{AB}$ in Eqs.(4) to perform the above experiment, we will obtain another set of B possible states: 
\beqa
  \rho_{B1}\,=\,\vert c_1\vert^2 \vert B_1\rangle\langle B_1\vert + c_1 c_2^{*}\vert B_1\rangle\langle B_2\vert
  + c_2 c_1^{*}\vert B_2\rangle\langle B_1\vert +\vert c_2\vert^2 \vert B_2\rangle\langle B_2\vert \,, \nonumber \\
  \rho_{B2}\,=\,\vert c_1\vert^2 \vert B_1\rangle\langle B_1\vert - c_1 c_2^{*}\vert B_1\rangle\langle B_2\vert
  - c_2 c_1^{*}\vert B_2\rangle\langle B_1\vert +\vert c_2\vert^2 \vert B_2\rangle\langle B_2\vert \,, \nonumber \\
  \rho_{B3}\,=\,\vert c_2\vert^2 \vert B_1\rangle\langle B_1\vert + c_2 c_1^{*}\vert B_1\rangle\langle B_2\vert
  + c_1 c_2^{*}\vert B_2\rangle\langle B_1\vert + \vert c_1\vert^2 \vert B_2\rangle\langle B_2\vert \,, \nonumber \\
  \rho_{B4}\,=\,\vert c_2\vert^2 \vert B_1\rangle\langle B_1\vert - c_2 c_1^{*}\vert B_1\rangle\langle B_2\vert
  - c_1 c_2^{*}\vert B_2\rangle\langle B_1\vert + \vert c_1\vert^2 \vert B_2\rangle\langle B_2\vert \,.
\eeqa
Compare this result with Eqs.(6) we find the B final states in Eqs.(8) contain more information of C initial state: the relative phase of the expanding coefficients $c_1$ and $c_2$. So we can conclude that the quantum channel capacity of $\vert \psi_{AB}\rangle\langle \psi_{AB}\vert$ in Eq.(7) is larger than that of $\rho_{AB}$ in Eqs.(4). However we cannot conclude that the channel $\rho_{AB}$ in Eqs.(4) doesn't have the ability to transport quantum information (the module squares of $c_1$ and $c_2$ are also quantum informations). Thus $\rho_{AB}$ in Eqs.(4) must contain a certain kind of quantum entanglement between A and B systems, though this kind of quantum entanglement isn't same as the quantum entanglement in Eq.(7). 

In general, we discuss the quantum entanglement contained in the states described by density matrices. We point out that density matrices can only be used to describe quantum states, thus the entanglement contained in density matrices is just quantum entanglement. Under this point most of the quantum separable states people considered before, such as the bipartite state in Eq.(1), are in fact quantum entangled states. 

\vspace{5mm} {\bf \Large Acknowledgments} \vspace{2mm}

The author thanks Prof. Hong-bo Zhu and Doctor Pei-lin Lang for the useful discussions with them.

\vspace{5mm} {\bf \Large Appendix} \vspace{2mm}

In this appendix we present the works of trying to quantify the quantum entanglement degrees of the states described by density matrices. It is obvious that the quantum entanglement degree in Eq.(7) is larger than that in Eqs.(4). We believe the increase of the quantum entanglement degree of Eq.(7) comes from the two terms $\half\vert A_1\rangle\langle A_2\vert B_1 \rangle\langle B_2\vert +\half\vert A_2\rangle\langle A_1\vert B_2 \rangle\langle B_1\vert$ which is the excess part of $\vert \psi_{AB}\rangle\langle \psi_{AB}\vert$ than $\rho_{AB}$. In general it can be concluded that the quantum entanglement degree of A and B systems can be determined by the expanding coefficients of A and B system's (reduced) density matrix. If A system's orthonormal bases are $\{ \vert A_i \rangle , i=1...n\}$, B system's orthonormal bases are $\{ \vert B_i \rangle , i=1...m\}$,  expand the density matrix $\rho_{AB}$ according to A system's orthonormal bases: 
\beq
  \rho_{AB}\,=\,\sum_{i,i^{\prime}=1}^n \rho_{ii^{\prime}}\vert A_i\rangle\langle A_{i^{\prime}}\vert\,,
\eeq
where the expanding coefficients $\{ \rho_{ii^{\prime}} \}$ can be written as follows:
\beq
  \rho_{ii^{\prime}}\,=\,\sum_{j,j^{\prime}=1}^m p_{ii^{\prime},jj^{\prime}} \vert B_j\rangle\langle B_{j^{\prime}} \vert\,.
\eeq
If all of $\{ \rho_{ii^{\prime}} \}$ are same, A and B systems will not get entangled with each other. So the more different $\{ \rho_{ii^{\prime}} \}$ are, the more large the quantum entanglement degree is. The difference between $\{ \rho_{ii^{\prime}} \}$ can be determined by the following quantity
\beq
  X_{i_1 i_1^{\prime}, i_2 i_2^{\prime}}\,=\,tr( \rho_{i_1 i_1^{\prime}} \rho_{i_2 i_2^{\prime}}^{\dagger} )
  \,=\,\sum_{j,j^{\prime}}^m p_{i_1 i_1^{\prime}, jj^{\prime}} p^{*}_{i_2 i_2^{\prime}, jj^{\prime}}\,.
\eeq
The matrix $X$ is a Hermitian matrix, so it can be diagonalized by an unitary matrix. The rank of matrix $X$ is just the number of the orthonormal bases of $\{ \rho_{ii^{\prime}} \}$. Obviously the rank of matrix $X$ is the bigger the better. But it is difficult to use matrix $X$ to quantify quantum entanglement degree, because the values of the matrix elements of $X$ also reflect the difference between pure states and mixed states (the trace of the module square of pure state is 1, but that of mixed state is less than 1) which is nothing to do with quantum entanglement. So up to the present we still haven't found a suitable method to quantify the quantum entanglement degree of a general quantum state.


\begin{thebibliography}{99}

\bibitem{c1}
A. Einstein, B. Podolsky and N. Rosen, Phys. Rev. 47 (1935) 777.

\bibitem{c2}
E. Schrodinger, Naturwissenschaften 23 (1935) 807.

\bibitem{c3}
J.S. Bell, Physics (N.Y.) 1 (1964) 195.

\bibitem{c4}
C.H. Bennett, G. Brassard, C. Crepeau, R. Jozsa, A. Peres, and W.K. Wootters, Phys. Rev. Lett. 70 (1993) 1895; \\
S. Albeverio and S.M. Fei, Phys. Lett. A 276 (2000) 8; \\
G.M. D¡¯Ariano, P.Lo Presti, M.F. Sacchi, Phys. Lett. A 272 (2000) 32; \\
S. Albeverio and S.M. Fei and W.L. Yang, Phys. Rev. A 66 (2002) 012301.

\bibitem{c5}
D. Bouwmeester, J.-W. Pan, K. Mattle, M. Elbl, H. Weinfurter and A. Zeilinger, Nature (London) 390 (1997) 575; \\
M. A. Nielsen, E. Knill and R. Laflamme, Nature 396 (1998) 52; \\
Jian-Wei Pan, Matthew Daniell, Sara Gasparoni, Gregor Weihs, Anton Zeilinger, Phys. Rev. Lett. 86 (2001) 4435; \\
Thomas Jennewein, Gregor Weihs, Jian-Wei Pan, Anton Zeilinger, Phys. Rev. Lett. 88 (2001) 017903; \\
Qiang Zhang, Alexander Goebel, Claudia Wagenknecht, Yu-Ao Chen, Bo Zhao, Tao Yang, Alois Mair, Joerg Schmiedmayer, Jian-Wei Pan, Nature Physics 2 (2006) 678.

\bibitem{c6}
A. K. Ekert, Phys. Rev. Lett. 67 (1991) 661; \\
D. Deutsch, A. K. Ekert, R. Jozsa, C. Macchiavello, S. Popescu and A. Sanpera, Phys. Rev. Lett. 77 (1996) 2818; \\
C.A. Fuchs, N. Gisin, R.B. Griffiths, C-S. Niu, and A. Peres, Phys. Rev. A 56 (1997) 1163.

\bibitem{c7}
C.H. Bennett and S.J. Wiesner, Phys. Rev. Lett. 69 (1992) 2881.

\bibitem{c8}
D.P. DiVincenzo, Science 270 (1995) 255; \\
M.A. Nielsen and I.L. Chuang, Quantum Computation and Quantum Information, Cambridge University Press, Cambridge, 2000.

\bibitem{c9}
A. Peres, ¡°Quantum Theory: Concepts and Methods¡±, Kluwer Academic Publishers (1995).

\bibitem{c10}
C. H. Bennett, D. P. DiVincenzo, J. A. Smolin, and W. K. Wootters, Phys. Rev. A 54 (1996) 3824; \\
V. Vedral, M.B. Plenio, M.A. Rippin, P. L. Knight, Phys. Rev. Lett. 78 (1997) 2275; \\
C. H. Bennett, D. P. DiVincenzo, T. Mor, P. W. Shor, J. A. Smolin and B. M. Terhal,
Phys. Rev. Lett. 82 (1999) 5385; \\
W. Dur, J. I. Cirac and R. Tarrach, Phys. Rev. Lett. 83 (1999) 3562; \\
P. Horodecki, M. Lewenstein, G. Vidal and I. Cirac, Phys. Rev. A 62 (2000) 032310; \\
S. Karnas and M. Lewenstein, Phys. Rev. A 64 (2001) 042313; \\
K. Chen and L.A. Wu, Phys. Lett. A 306 (2002) 14; \\
S.M. Fei, J. Jost, X.Q. Li-Jost and G.F. Wang, Phys. Lett. A 310 (2003) 333; \\
O. Rudolph, Phys. Rev. A 67 (2003) 032312.

\bibitem{c11}
R. F. Werner, Phys. Rev. A 40 (1989) 4277.

\bibitem{c12}
Asher Peres, Phys. Rev. Lett. 77 (1996) 1413; \\
M. Horodecki, P. Horodecki and R. Horodecki, Springer Tracts in Mod. Phy. 173 (2001) 151; \\
B.M. Terhal, Theor. Comput. Sci. 287 (2002) 313; \\
K. Chen, S. Albeverio and S.M. Fei, Phys. Rev. Lett. 95 (2005) 210501.

\end{thebibliography}
\end{document}